\PassOptionsToPackage{unicode}{hyperref}
\PassOptionsToPackage{hyphens}{url}
\documentclass[
  11pt,
]{article}
\usepackage{amsmath,amssymb}
\usepackage{iftex}
\ifPDFTeX
  \usepackage[T1]{fontenc}
  \usepackage[utf8]{inputenc}
  \usepackage{textcomp} 
\else 
  \usepackage{unicode-math} 
  \defaultfontfeatures{Scale=MatchLowercase}
  \defaultfontfeatures[\rmfamily]{Ligatures=TeX,Scale=1}
\fi
\ifPDFTeX\else
\fi
\IfFileExists{upquote.sty}{\usepackage{upquote}}{}
\IfFileExists{microtype.sty}{
  \usepackage[]{microtype}
  \UseMicrotypeSet[protrusion]{basicmath} 
}{}
\usepackage{xcolor}
\usepackage[margin=1in]{geometry}
\usepackage{longtable,booktabs,array}
\usepackage{calc} 
\usepackage{etoolbox}
\makeatletter
\patchcmd\longtable{\par}{\if@noskipsec\mbox{}\fi\par}{}{}
\makeatother
\IfFileExists{footnotehyper.sty}{\usepackage{footnotehyper}}{\usepackage{footnote}}
\makesavenoteenv{longtable}
\usepackage{graphicx}
\makeatletter
\def\maxwidth{\ifdim\Gin@nat@width>\linewidth\linewidth\else\Gin@nat@width\fi}
\def\maxheight{\ifdim\Gin@nat@height>\textheight\textheight\else\Gin@nat@height\fi}
\makeatother
\setkeys{Gin}{width=\maxwidth,height=\maxheight,keepaspectratio}
\makeatletter
\def\fps@figure{htbp}
\makeatother
\usepackage{pos}
\usepackage[absolute,overlay]{textpos}

\author*[a,b]{Sofie Martins}
\author[a]{Erik Kjellgren}
\author[a]{Emiliano Molinaro}
\author[a,b]{Claudio Pica} 
\author[a,b]{Antonio~Rago}

\affiliation[a]{University of Southern Denmark, Campusvej 55, 5230 Odense M, Denmark}
\affiliation[b]{$\hbar$QTC, University of Southern Denmark, Campusvej 55, 5230 Odense M, Denmark}

\emailAdd{martinss@imada.sdu.dk}

\abstract{We are improving one of the available lattice software packages \texttt{HiRep} by adding GPU acceleration supporting highly-optimized simulations on both NVIDIA and AMD GPUs. \texttt{HiRep} allows lattice simulations of theories with fermions in higher representations and a variable number of colors in the gauge group. The development is accompanied by an overall software quality improvement in the build system, testing, and documentation, adding features for both CPUs and GPUs. The software is available under \texttt{https://github.com/claudiopica/HiRep}.}

\FullConference{%
European network for Particle physics, Lattice field theory and Extreme computing (EuroPLEx2023)\\
11-15 September 2023\\
Berlin, Germany\\}

\usepackage{flafter}
\usepackage{float}
\usepackage{pos}
\usepackage{natbib}
\usepackage{subfigure}
\usepackage{booktabs}
\usepackage{longtable}
\usepackage{array}
\usepackage{multirow}
\usepackage{wrapfig}
\usepackage{colortbl}
\usepackage{pdflscape}
\usepackage{tabu}
\usepackage{threeparttable}
\usepackage{threeparttablex}
\usepackage[normalem]{ulem}
\usepackage{makecell}
\usepackage{xcolor}
\ifLuaTeX
  \usepackage{selnolig}  
\fi
\usepackage[]{natbib}
\nocite{*}
\usepackage{bookmark}
\IfFileExists{xurl.sty}{\usepackage{xurl}}{} 
\hypersetup{
  pdftitle={GPU-accelerated Higher Representations of Wilson Fermions with HiRep},
  hidelinks,
  pdfcreator={LaTeX via pandoc}}

\title{GPU-accelerated Higher Representations of Wilson Fermions with HiRep}
\date{\vspace{-2.5em}}

\begin{document}
\maketitle

{
\setcounter{tocdepth}{2}
\tableofcontents
}
\section{Motivation}\label{motivation}

Predictions from models of physics beyond the standard model often depend on input from strongly coupled gauge theories of variable numbers of colors and fermion representations. Simulations on the lattice can quickly become computationally expensive. Using Wilson Fermions constitutes, to this day, one of the cheapest options to explore theories with different gauge groups. In particular, the simulation of theories with higher dimensional fermion representations at sufficient precision requires generally a large amount of computational resources.

The lattice library \texttt{HiRep} allows simulations of higher representations of Wilson fermions with general gauge groups. In the following we present results for the porting of the library to state-of-the-art GPU accelerators. The software is available at

\vspace{0.5cm}
\centering

\texttt{https://github.com/claudiopica/HiRep} \flushleft

\section{Software Quality}\label{software-quality}

We improved the quality assurance of the software by introducing automated unit tests, testing different numbers of colors and fermion representations, MPI support, OpenMP support, GPU support, and clover improvement. for each commit and code coverage reporting with \texttt{codecov} supplying the user with detailed information about which lines of the code are tested. To ensure the extensibility of the library, we generate documentation \citep{hirepdocs} including a user manual, developer handbook and \texttt{Doxygen} function reference and improve code readability using a code formatting check.

For a fast and simple compilation for different gauge groups and fermion representations, we have updated the build system to \texttt{ninja-build} \citep{ninja}, a lower-level build system that avoids large compilation times. The library is largely independent of external modules or libraries, requiring \texttt{gcc}, \texttt{perl} and \texttt{ninja-build} and optionally \texttt{CUDA} (GPU acceleration), \texttt{MPI} (parallelization) and \texttt{hwloc} (hardware topology). Configuration of build setup and compilation variables is supplied over a build help text.

New benchmarking code has been developed for gaining quick information on the configuration of the software on a new cluster or supercomputer.

\subsection{Memory Access Patterns}\label{memory-access-patterns}

Lattice simulations are usually bound by the speed of the memory access. While modern CPUs can access memory with high flexibility, access patterns are crucial to optimal GPU performance. GPU manufacturers cite here a peak value of global memory access, which can be reached by reading memory block after block using sequential GPU threads. However, an ideal memory arrangement for the GPU is not usually the most intuitive way to store lattice data.

\begin{figure}

{\centering \includegraphics{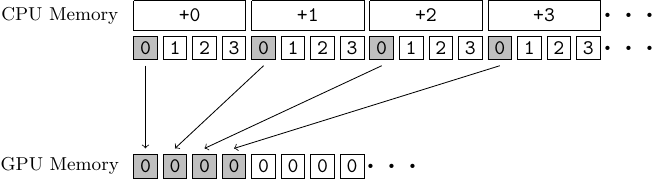} 

}

\caption{Comparison of memory access patterns in CPU and GPU simulations}\label{fig:memory-coalescing}
\end{figure}

For simulations on the CPU, it is sufficient to store memory site-by-site as in figure \ref{fig:memory-coalescing}. However, this is a problem for reading memory from the GPU because the kernels will only need to read the first component or first real number of the site and then move on to the other components. Therefore, the ideal memory access pattern is to store blocks of components as illustrated in figure \ref{fig:memory-coalescing}. Instead of components, we store the data in the smallest memory units we read, single real numbers of single or double precision.

Unfortunately, this memory striding is incompatible with the legacy geometry of \texttt{HiRep}. In the legacy geometry, we categorize sites of the local lattices into sites that, during the hopping term application, do not depend on communications (\emph{bulk}), sites depending on communications (\emph{boundary}), and sites that the boundary depends on (\emph{halo} or \emph{receive buffers}) and arrange the sites of the four-dimensional lattice in such a way that the boundary and receive buffers are contiguous in memory. This way, we can identify the contiguous memory region and pass it to MPI.

On a four-dimensional lattice, this is rarely trivially possible. The simplest solution to this problem is to collect the sites that need to be sent over MPI into a separate send buffer, copying all sites on the boundary of the inner lattice. The old geometry aims to minimize the number of copies by adding duplicated sites in place and synchronizing them. Unfortunately, in combination with the striding on the GPU, this is impossible because the striding breaks the buffers into non-contiguous pieces.

\begin{figure}
\begin{minipage}[c]{0.5\linewidth}
\centering

\begin{center}\includegraphics{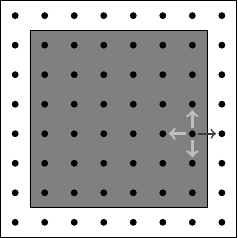} \end{center}

\end{minipage}\begin{minipage}[c]{0.5\linewidth}
\centering

\begin{center}\includegraphics{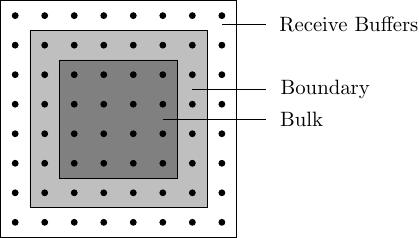} \end{center}

\end{minipage}
\caption{New geometry (left) and legacy geometry (right), illustrated on 2D lattices.}
\label{fig-test}
\end{figure}

Due to this, we implemented the new geometry using a complete copy to a send buffer. This geometry has only two categories of sites: bulk and receive buffers. Instead of waiting for communications before we compute the sites on the boundary, we only mask the directions in the hopping term that can only be executed after the communications are complete. This has the advantage that more time is spent in the bulk computation, giving us more time to mask communications. Find illustrations in figure \ref{fig-test}, based on figures published in \citep{hirepdocs}.

\subsection{Send buffer synchronization}\label{send-buffer-synchronization}

The new geometry has the disadvantage that the send and receive buffers are no longer contiguous. Consequently, we need to synchronize the non-contiguous pieces of information to a contiguous send buffer. This synchronization is slightly more complicated than the synchronization on the legacy geometry; however, the execution times of both operations are negligible compared to the bulk calculations of the Dirac operator applications on typical local lattice sizes.

\begin{figure}

{\centering \includegraphics{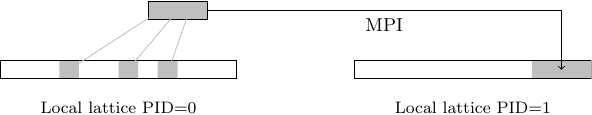} 

}

\caption{Illustration of the send buffer synchronization in the new geometry}\label{fig:sendbuffer-sync}
\end{figure}

We are speeding up communications by partially applying the Dirac operator before the synchronization to the send buffer. As a result, it is only necessary to send a half-spinor, cutting the communication times substantially, see for example \citep{openqcd}.

\section{Features}\label{features}

\subsection{Wilson-Dirac Operator}\label{wilson-dirac-operator}

The GPU-ported version of \texttt{HiRep} supports a GPU-accelerated Wilson-Dirac operator in single and double precision for general higher representations of SU(\(N_c\)) for any number of colors. We are supporting even-odd preconditioning, clover \citep{Sheikholeslami:1985ij}, exponentiated clover improvement \citep{Francis:2019muy}, and the Lüscher-Weisz gauge action \citep{Luscher:1984xn}. Available are not only the HMC but also the RHMC \citep{Kennedy:1998cu, Clark:2003na}, Hasenbusch acceleration \citep{Hasenbusch:2001ne} and a selection of integrators: Leapfrog, 2nd and 4th order Omelyan integrators \citep{OMELYAN2003272}.

\subsubsection{Strong and weak scaling}\label{strong-and-weak-scaling}

\begin{figure}

{\centering \includegraphics{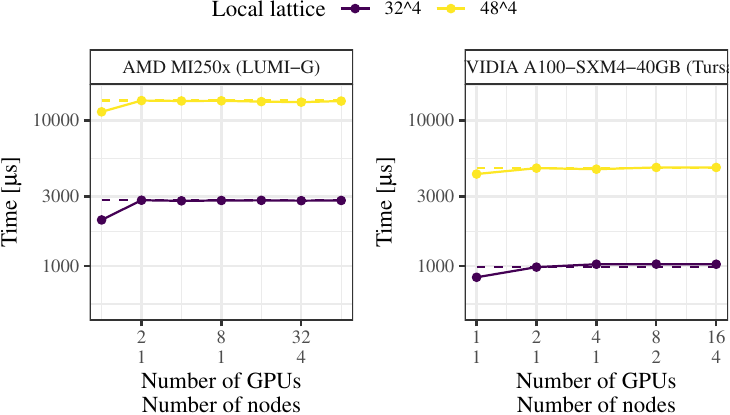} 

}

\caption{Weak scaling of time spent on a single application of the Wilson-Dirac operator for an SU(2) gauge group with fermions in the fundamental representation.}\label{fig:ws-su2}
\end{figure}

\begin{figure}

{\centering \includegraphics{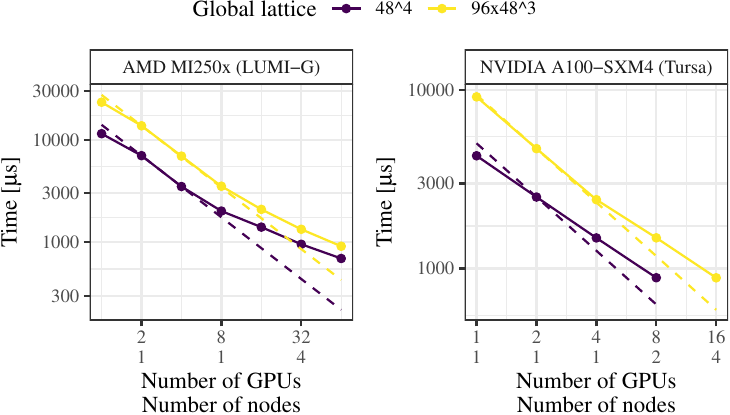} 

}

\caption{Strong scaling of time spent on a single application of the Wilson-Dirac operator for an SU(2) gauge group with fermions in the fundamental representation.}\label{fig:ss-su2}
\end{figure}

We observe almost ideal weak scaling on larger clusters, even for multi-node simulations. We show in figure \ref{fig:ws-su2} the weak scaling of the smallest possible kernel: An SU(2) gauge group with fermions in the fundamental representation. Both on LUMI-G and Tursa we see that the computations are masking the communications for single and multi-node simulations.

The strong scaling in figure \ref{fig:ss-su2}, however, is showing less ideal behavior. Both on Tursa and LUMI-G the loss of efficiency is mainly caused by a strong dependence of the kernel execution time on the total memory movement. In a region where the local lattices become small, the execution time is dominated by the kernel call overheads and ideal scaling is not expected.

\subsubsection{\texorpdfstring{Large-\(N_c\) scaling}{Large-N\_c scaling}}\label{large-n_c-scaling}

A crucial property of the library is the ability to use fermions in arbitrary higher representations. For the CPU version the operation on the fields are site-wise operations. As a result the computational throughput per thread or process is increasing with the square of the dimension of the fermion representation \(N_{\mathrm{f}}^2\), the amount of computations in the matrix multiplication of the represented gauge field with the spinor vector components in the Dirac operator. Additionally, the amount of memory each thread or process is reading increases, since the amount of data stored in a single spinor and a single SU(\(N_{\mathrm{f}}\)) matrix increases with \(\sim N_{\mathrm{f}}\) and \(\sim N_{\mathrm{f}}^2\) respectively.

While on the CPU we pay a performance penalty proportional to the additional computation and memory, the performance on the GPU can degrade very quickly and become completely unfeasible due to register spilling: In contrast to CPUs GPUs do not have an L3 cache and if too much memory is used in the local thread, it `spills' directly to global memory. Accessing global memory, however, is substantially slower.

In order to allow scaling of the code to large fermion representations, we have to parallelize in the fermion dimension. However, this parallelization is non-trivial due to the matrix multiplication. Correspondingly, a large-\(N_{\mathrm{f}}\)-improved kernel might not perform better than the naive implementation for small representations such as the ones of QCD-like theories. For this, we implemented two versions of this kernel, one is parallelized. It can be activated by changing the compilation variables, which is implemented by passing macro definitions to the preprocessor. The macro \texttt{LARGE\_N} switches to the parallelized kernel. However, on some cards the non-parallelized kernel still performs better for QCD-like theories. As a result, the default is the non-parallelized kernel.

\begin{figure}

{\centering \includegraphics{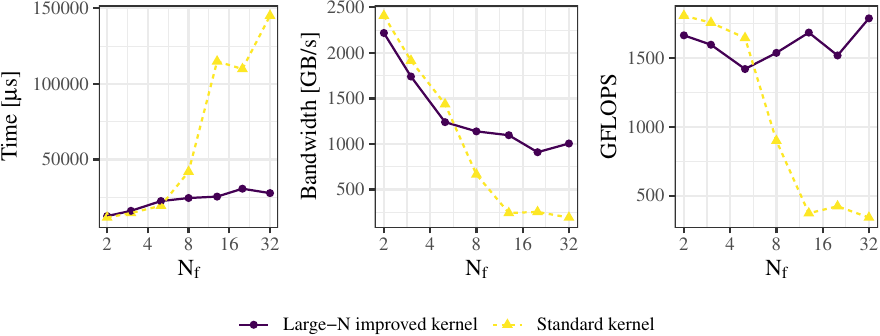} 

}

\caption{Large-$N_c$ scaling of performance metrics of applications of the Wilson-Dirac operator with an SU($N_c$) gauge group and fermions in the fundamental representation.}\label{fig:large-N}
\end{figure}

Figure \ref{fig:large-N} shows the execution time, bandwidth, and computational throughput as a function of the dimension of the fermion representation \(N_\mathrm{f}\). To avoid effects due to different memory sizes, we scaled the lattice size down when increasing \(N_\mathrm{f}\) to keep the total amount of memory moved constant. The values given are interpolated since this was impossible due to geometric constraints.

The figures show a steep increase in execution time for the unimproved kernel for large gauge groups. While the scaling in time for the improved kernel is not ideal, it is more stable. The computational throughput of the kernel is largely constant. The drop in bandwidth can be explained by the fact that the kernel becomes compute-bound for larger \(N_{\mathrm{f}}\). Still, the peak theoretical occupancy is limited because the kernel is not particularly lightweight and needs many registers. Consequently, the card is exhausting its capabilities in terms of computation for all \(N_\mathrm{f}\). This additionally causes a drop in the effective bandwidth.

\subsection{Linear Algebra}\label{linear-algebra}

An efficient inversion of the Wilson-Dirac operator further depends on efficient linear algebra operations, particularly for the spinors. The single and double-precision implementations of all fields available in the library are easily extendable, simplify use, and reach 80-90\% of the theoretical peak performance on NVIDIA GPUs.

\begin{figure}

{\centering \includegraphics{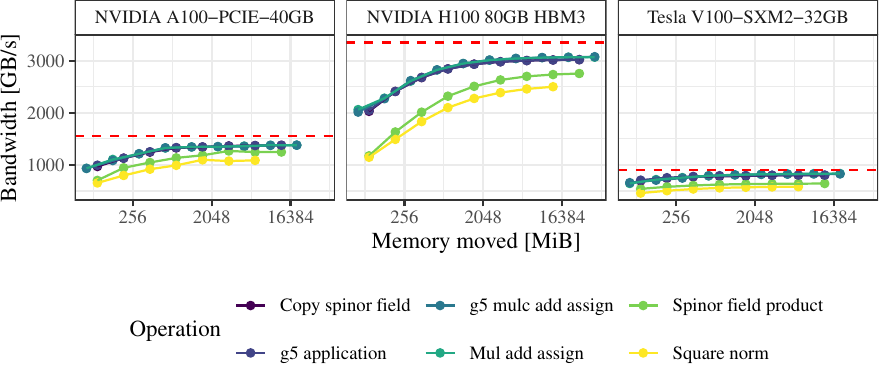} 

}

\caption{Bandwidths reached for linear algebra operations on different types of NVIDIA GPUs. Tested for an SU(3) gauge group with fermions in the fundamental representation.}\label{fig:lina}
\end{figure}

In Figure \ref{fig:lina}, we test the achieved bandwidth of selected linear algebra operations on a single device for different NVIDIA GPUs. We find that approximately 90\% of the peak performance is reached for simple operations such as the identity operation (\emph{Copy spinor field}) or a \(\gamma_{5}\) application (\emph{g5 application}) for sufficiently high memory movement. We see that there is a dependence on the amount of data we moved since the kernels have a call overhead and can only reach peak values if the GPUs have sufficient work to do for this overhead to be small. More intensive operations such as a multiply-assign with a complex number and application of \(\gamma_5\) (\emph{g5 mulc add assign}) are also saturating capabilities of the device.

Another critical operation type are \emph{reductions}, which map the field to a single number, such as inner products and norms. These are implemented to take the site-wise product or norms and reduce the result. Consequently, they consist of multiple kernels in sequence, each needing to read a substantial amount of data. It is expected that these operations do not saturate the theoretical peak because it accounts only for a first single read. However, the necessity of calling two kernels causes two stages where global memory needs to be accessed.

While the site-wise operations are kernels that are part of the \texttt{HiRep} library, the reductions are offloaded to \texttt{cub}, a library that has been part of the CUDA SDK since CUDA 11.

\subsection{Inverters}\label{inverters}

\texttt{HiRep} supports the Conjugate Gradient, stabilized Bi-Conjugate Gradient \citep{doi:10.1137/0913035, doi:10.1137/0914062}, and QMR\(\gamma_5\) inverters, which are ported to GPUs. In the future, we are planning to implement domain decomposition methods, such as in \citep{Luscher:2003qa, Babich:2010qb, Frommer:2013fsa} that reduce the amount of communications during the inversion additionally, which can be useful for small local lattices and ill-conditioned operators.

\subsection{HMC}\label{hmc}

Configuration generation with the HMC is supported for GPUs. The following algorithmic checks are done following and checking consistency with \citep{DelDebbio:2008zf} using an SU(2) gauge group with two fermions in the adjoint representation, using two RHMC monomials at \(\beta=2.0\) and \(m=-1.34\), both on an NVIDIA V100 GPU (via UCloud DeiC Interactive HPC) and, in addition, for the HIP-ported version tested on AMD MI250x GPUs (LUMI-G).

\begin{figure}

{\centering \includegraphics{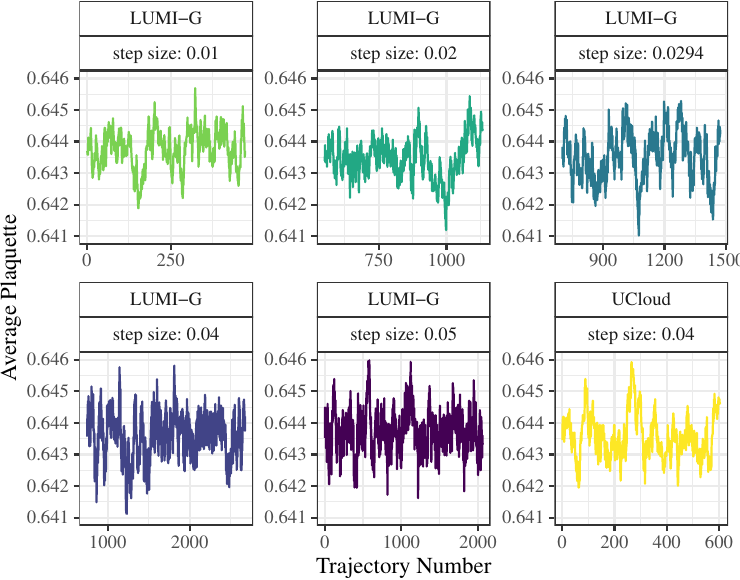} 

}

\caption{Average plaquette of ensembles}\label{fig:unnamed-chunk-6}
\end{figure}

\begin{figure}

{\centering \includegraphics{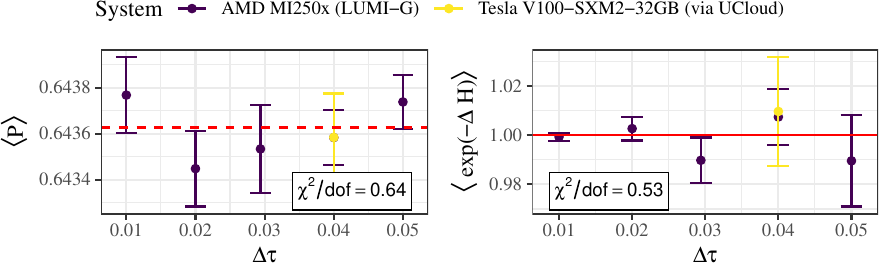} 

}

\caption{Check that the average plaquette value does not depend on the step size in the integrator (left). Check that the simulations obey the Creutz equality (right).}\label{fig:unnamed-chunk-10}
\end{figure}

The determined value of the average plaquette is independent of the step size in the molecular dynamics integration. We further check the Creutz equality, see \citep{PhysRevD.38.1228}, stating that \begin{equation}
\langle\mathrm{exp}(-\Delta H)\rangle = 1
\end{equation}

While we do see, as expected, larger uncertainties for larger step sizes, the ensembles show results consistent with this equality.

\begin{figure}

{\centering \includegraphics{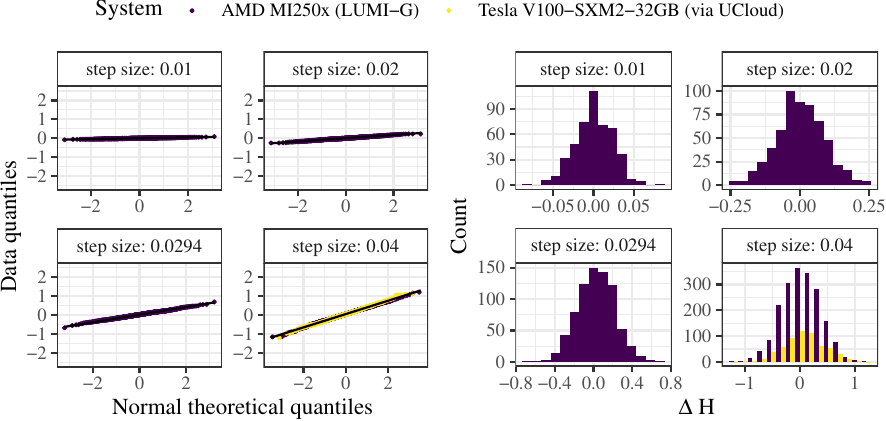} 

}

\caption{Distribution of Hamiltonian violations for selected step sizes}\label{fig:unnamed-chunk-13}
\end{figure}

We further examine the distribution of Hamiltonian violations of the HMC trajectories. For sufficiently small step sizes, the distribution is normal, as expected, which is demonstrated both by the histograms on the right as well as the quantile-quantile plots on the left, which show the values of Hamiltonian violation on the y-axis and the expected quantile they should be located in if they were Gaussian on the x-axis. A perfectly normal distributed data vector forms a straight line in these plots. The slope of this line is the standard deviation of the data.

\begin{figure}

{\centering \includegraphics{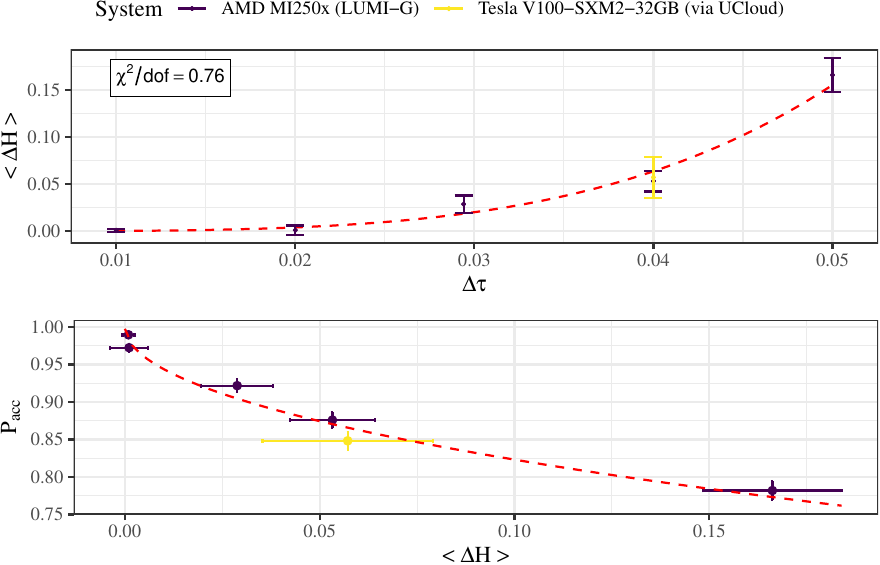} 

}

\caption{Check analytic behavior of acceptance rate (bottom) and scaling of the integrator with step size (top).}\label{fig:unnamed-chunk-15}
\end{figure}

Lastly, we check that the Hamiltonian violations of the 2nd-order Omelyan integrator \citep{OMELYAN2003272} are increasing as \begin{equation}
\Delta H \sim \Delta \tau^4, 
\end{equation} see \citep{TAKAISHI20006}, by fitting. Additionally, we check the asymptotic analytical relation from \citep{Gupta:1990ka} \begin{equation}
P_{\mathrm{acc}} \cong \mathrm{erfc}\left(\sqrt{\dfrac{\langle\Delta H\rangle}{2}}\right)
\end{equation} between the acceptance rate and Hamiltonian violations. We find that the relations are fulfilled and that the algorithm is working.

\section{Conclusion and outlook}\label{conclusion-and-outlook}

The ported code is efficient and scales well. Substantial improvements in overall software quality have been made. While configuration generation with the HMC is efficient, performance improvements can still be achieved using domain decomposition methods, as they minimize the need for communications.

\section{Acknowledgements}\label{acknowledgements}

This project has received funding from the European Union's Horizon 2020 research and innovation program under the Marie Sk\l odowska-Curie grant agreement \textnumero 813942. Testing, development, and benchmarking of this software was possible using resources on LUMI-G provided by the Danish eInfrastructure Consortium under grant application number DeiC-SDU-N5-2024055 and NVIDIA V100, A100, and H100 nodes provided by the UCloud DeiC Interactive HPC system managed by the eScience Center at the University of Southern Denmark. We are grateful to Jacob Finkenrath and Martin Hansen for valuable discussions.

\renewcommand\refname{References}
  \bibliography{literature.bib}

\begin{thebibliography}{10}

\bibitem{ninja}
The ninja build system.
\newblock \url{https://ninja-build.org/}, 2013--2024.

\bibitem{Babich:2010qb}
R.~Babich, J.~Brannick, R.~C. Brower, M.~A. Clark, T.~A. Manteuffel, S.~F.
  McCormick, J.~C. Osborn, and C.~Rebbi.
\newblock {Adaptive multigrid algorithm for the lattice Wilson-Dirac operator}.
\newblock {\em Phys. Rev. Lett.}, 105:201602, 2010.

\bibitem{Clark:2003na}
M.~A. Clark and A.~D. Kennedy.
\newblock {The RHMC algorithm for two flavors of dynamical staggered fermions}.
\newblock {\em Nucl. Phys. B Proc. Suppl.}, 129:850--852, 2004.

\bibitem{PhysRevD.38.1228}
Michael Creutz.
\newblock Global monte carlo algorithms for many-fermion systems.
\newblock {\em Phys. Rev. D}, 38:1228--1238, Aug 1988.

\bibitem{DelDebbio:2008zf}
Luigi Del~Debbio, Agostino Patella, and Claudio Pica.
\newblock {Higher representations on the lattice: Numerical simulations. SU(2)
  with adjoint fermions}.
\newblock {\em Phys. Rev. D}, 81:094503, 2010.

\bibitem{hirepdocs}
Claudio~Pica et. al.
\newblock Documentation of hirep.
\newblock \url{https://claudiopica.github.io/HiRep/}.

\bibitem{openqcd}
Martin~Lüscher et. al.
\newblock Openqcd.
\newblock \url{https://luscher.web.cern.ch/luscher/openQCD/}.

\bibitem{Francis:2019muy}
Anthony Francis, Patrick Fritzsch, Martin L\"uscher, and Antonio Rago.
\newblock {Master-field simulations of O($a$)-improved lattice QCD: Algorithms,
  stability and exactness}.
\newblock {\em Comput. Phys. Commun.}, 255:107355, 2020.

\bibitem{Frommer:2013fsa}
Andreas Frommer, Karsten Kahl, Stefan Krieg, Bj\"orn Leder, and Matthias
  Rottmann.
\newblock {Adaptive Aggregation-Based Domain Decomposition Multigrid for the
  Lattice Wilson--Dirac Operator}.
\newblock {\em SIAM J. Sci. Comput.}, 36(4):A1581--A1608, 2014.

\bibitem{Gupta:1990ka}
Sourendu Gupta, A.~Irback, F.~Karsch, and B.~Petersson.
\newblock {The Acceptance Probability in the Hybrid Monte Carlo Method}.
\newblock {\em Phys. Lett. B}, 242:437--443, 1990.

\bibitem{doi:10.1137/0914062}
Martin~H. Gutknecht.
\newblock Variants of bicgstab for matrices with complex spectrum.
\newblock {\em SIAM Journal on Scientific Computing}, 14(5):1020--1033, 1993.

\bibitem{Hasenbusch:2001ne}
Martin Hasenbusch.
\newblock {Speeding up the hybrid Monte Carlo algorithm for dynamical
  fermions}.
\newblock {\em Phys. Lett. B}, 519:177--182, 2001.

\bibitem{Kennedy:1998cu}
A.~D. Kennedy, Ivan Horvath, and Stefan Sint.
\newblock {A New exact method for dynamical fermion computations with nonlocal
  actions}.
\newblock {\em Nucl. Phys. B Proc. Suppl.}, 73:834--836, 1999.

\bibitem{Luscher:1984xn}
M.~Luscher and P.~Weisz.
\newblock {On-shell improved lattice gauge theories}.
\newblock {\em Commun. Math. Phys.}, 98(3):433, 1985.
\newblock [Erratum: Commun.Math.Phys. 98, 433 (1985)].

\bibitem{Luscher:2003qa}
Martin Luscher.
\newblock {Solution of the Dirac equation in lattice QCD using a domain
  decomposition method}.
\newblock {\em Comput. Phys. Commun.}, 156:209--220, 2004.

\bibitem{OMELYAN2003272}
I.P. Omelyan, I.M. Mryglod, and R.~Folk.
\newblock Symplectic analytically integrable decomposition algorithms:
  classification, derivation, and application to molecular dynamics, quantum
  and celestial mechanics simulations.
\newblock {\em Computer Physics Communications}, 151(3):272--314, 2003.

\bibitem{Sheikholeslami:1985ij}
B.~Sheikholeslami and R.~Wohlert.
\newblock {Improved Continuum Limit Lattice Action for QCD with Wilson
  Fermions}.
\newblock {\em Nucl. Phys. B}, 259:572, 1985.

\bibitem{TAKAISHI20006}
Tetsuya Takaishi.
\newblock Choice of integrator in the hybrid monte carlo algorithm.
\newblock {\em Computer Physics Communications}, 133(1):6--17, 2000.

\bibitem{doi:10.1137/0913035}
H.~A. van~der Vorst.
\newblock Bi-cgstab: A fast and smoothly converging variant of bi-cg for the
  solution of nonsymmetric linear systems.
\newblock {\em SIAM Journal on Scientific and Statistical Computing},
  13(2):631--644, 1992.

\end{thebibliography}

\end{document}